\documentclass[12pt]{article}
\usepackage{amsmath}
\usepackage{amssymb}
\usepackage{amsfonts}
\usepackage{epsfig}

\begin{document}

\title{Stochastic solutions with derivatives and non-polynomial terms: The
scrape-off layer equations}
\author{R. Vilela Mendes \thanks{%
vilela@cii.fc.ul.pt, rvilela.mendes@gmail.com} \\
CMAF - Instituto de Investiga\c{c}\~{a}o Interdisciplinar UL \\
(Av. Gama Pinto 2, 1649-003, Lisbon)\\
Instituto de Plasmas e Fus\~{a}o Nuclear, IST \\
(Av. Rovisco Pais, 1049-001 Lisbon)}
\date{ }
\maketitle

\begin{abstract}
The construction of stochastic solutions for nonlinear partial differential
equations is a powerful method to obtain new exact results and to develop
efficient numerical algorithms, in particular when domain decomposition
techniques are used.

This paper deals with the problems that arise when the nonlinear terms are
nonpolynomial or involve derivatives. A set of equations of relevance for
plasma physics is used as a testing ground for these problems.
\end{abstract}

\section{Introduction}

A stochastic solution of a linear or nonlinear partial differential equation
is a stochastic process which, when started from a particular point in the
domain generates after a time $t$ a boundary measure which, integrated over
the initial condition at $t=0$, provides the solution at the point $x$ and
time $t$. For example for the heat equation%
\begin{equation}
\partial _{t}u(t,x)=\frac{1}{2}\frac{\partial ^{2}}{\partial x^{2}}%
u(t,x)\qquad \text{\textnormal{with}}\qquad u(0,x)=f(x)  \label{1.1}
\end{equation}%
the stochastic process is Brownian motion and the solution is 
\begin{equation}
u(t,x)=\mathbb{E}_{x}f(X_{t})  \label{1.2}
\end{equation}%
$\mathbb{E}_{x}$ meaning the expectation value, starting from $x$, of the
process%
\begin{equation}
dX_{t}=dB_{t}  \label{1.3}
\end{equation}%
The domain here is $\mathbb{R}\times \left[ 0,t\right) $ and the expectation
value in (\ref{1.2}) is the inner product $\left\langle \mu
_{t},f\right\rangle $ of the initial condition $f$ with the measure $\mu
_{t} $ generated by the Brownian motion at the $t-$boundary. An important
condition for the stochastic process (Brownian motion in this case) to be
considered \textit{the} solution of the equation is the fact that the same
process works for any initial condition. This should be contrasted with
stochastic processes constructed from particular solutions.

That the solutions of linear elliptic and parabolic equations, both with
Cauchy and Dirichlet boundary conditions, have a probabilistic
interpretation is a classical result and a standard tool in potential theory 
\cite{Getoor} \cite{Bass1} \cite{Bass2}. In contrast with the linear
problems, explicit solutions in terms of elementary functions or integrals
for nonlinear partial differential equations are only known in very
particular cases. Therefore the construction of solutions through stochastic
processes, for nonlinear equations, has become an active field in recent
years. The first stochastic solution for a nonlinear PDE was constructed by
McKean \cite{McKean} for the KPP equation. Later on, the exit measures
provided by diffusion plus branching processes \cite{Dynkin1} \cite{Dynkin2}
as well as the stochastic representations recently constructed for the
Navier-Stokes \cite{Jan} \cite{Waymire} \cite{Bhatta1} \cite{Ossiander} \cite%
{Orum}, the Vlasov-Poisson \cite{Vilela1} \cite{Vilela2} \cite{Vilela4}, the
Euler \cite{Vilela3} and a fractional version of the KPP equation \cite%
{Cipriano} define solution-independent processes for which the mean values
of some functionals are solutions to these equations. Therefore, they are 
\textit{exact stochastic solutions}.

In the stochastic solutions one deals with a process that starts from the
point where the solution is to be found, a functional being then computed
when the process reaches the boundary. In addition to providing new exact
results, the stochastic solutions are also, in some cases, a promising tool
for numerical implementation. This is because stochastic simulation only
grows with the dimension of the process, whereas a deterministic algorithm
grows exponentially with the dimension of the space. In addition, because of
the independence of the sample paths of the process, they are a natural
choice for parallel and distributed computation.

Stochastic algorithms are also used for domain decomposition purposes \cite%
{Acebron1} \cite{Acebron2} \cite{Acebron3}. One decomposes the space in
subdomains and then uses in each one a deterministic algorithm with
Dirichlet boundary conditions, the values on the boundaries being determined
by a stochastic algorithm, thus minimizing the time-consuming communication
problem between domains.

There are basically two methods to construct stochastic solutions. The first
method, which will be called the McKean method, is essentially a
probabilistic interpretation of the Picard series. The differential equation
is written as an integral equation which is rearranged in a such a way that
the coefficients of the successive terms in the Picard iteration obey a
normalization condition. The Picard iteration is then interpreted as an
evolution and branching process, the stochastic solution being equivalent to
importance sampling of the normalized Picard series. The second method
constructs the boundary measures of a measure-valued stochastic process (a
superprocess) and obtain the solutions of the differential equation by a
scaling procedure. For a comparison of the two methods refer to \cite%
{Vilela5}.

To extend the construction of stochastic solutions to cases more general
than those dealt with in the past, techniques must be developed to handle
derivatives and nonpolynomial interactions\footnote{%
Here one is concerned with the construction of stochastic solutions using
McKean's method. Notice that the construction of Dynkin's superprocesses is
also restricted to nonlinear terms $u^{\alpha }$ with $0<\alpha \leq 2$. A
plausible conjecture is that, to extend the application of superprocesses to
more general non-linear equations, one should move from processes on
measures to processes on general distributions.}. Sometimes the direct
handling of derivatives may be avoided if the derivative of the propagation
kernel is smooth. This is the case in the configuration space Navier-Stokes
equation \cite{Ossiander}, where by an integration by parts the derivative
of the heat kernel is controlled by a majorizing kernel and absorbed in the
probability measure. However, in general, this is not possible.

In this paper, the construction of stochastic solutions, for differential
equations involving derivatives and nonpolynomial interactions, will be
carried out for two systems of equations which describe plasma turbulence in
the scrape-off layer. In both cases we deal with the Cauchy problem, 
namely the equations are defined in the full space with initial conditions at
$t=0$. This is the most natural setting when the McKean approach is used. 
Spatial boundary conditions are easier to implement through the superprocess
formulation, with or without a scaling limit (see \cite{Vilela5}).

\section{A system of scrape-off layer equations (SOLEDGE 2D)}

The SOLEDGE-2D\ equations are \cite{Bufferand}%
\begin{eqnarray}
\partial _{t}N+\frac{1}{q}\partial _{\theta }\Gamma +\frac{\chi }{\eta }N
&=&D\partial _{r}^{2}N  \notag \\
\partial _{t}\Gamma +\frac{1}{q}\left( 1-\chi \right) \partial _{\theta
}\left( \frac{\Gamma ^{2}}{N}+N\right) +\frac{\chi }{\eta }\left( \Gamma
-\Gamma _{0}\right) &=&\nu \partial _{r}^{2}\Gamma  \label{2.0}
\end{eqnarray}%
where $\Gamma $ and $N$ are the dimensionless parallel momentum and density, 
$\left( r,\theta \right) $ are the radial and poloidal coordinates and the
mask function $\chi $ equals one in a region where an obstacle is located
and zero elsewhere.

To construct a stochastic representation for the solution one needs to
identify a stochastic process associated to the linear component (to the
full linear component or part of it) and then, through an integral equation,
construct the branching mechanism representing the nonlinear part.

\subsection{The linear part, $\protect\chi =0$}

The linear part of the system for $\chi =0$ is:%
\begin{eqnarray}
\partial _{t}N+\frac{1}{q}\partial _{\theta }\Gamma &=&D\partial _{r}^{2}N 
\notag \\
\partial _{t}\Gamma +\frac{1}{q}\partial _{\theta }N &=&\nu \partial
_{r}^{2}\Gamma  \label{2.1}
\end{eqnarray}%
Given the initial conditions at time zero $\left( 
\begin{array}{l}
N\left( 0,r,\theta \right) \\ 
\Gamma \left( 0,r,\theta \right)%
\end{array}%
\right) $ the solution of this system is%
\begin{equation}
\left( 
\begin{array}{l}
N\left( t,r,\theta \right) \\ 
\Gamma \left( t,r,\theta \right)%
\end{array}%
\right) =\exp t\left\{ -\frac{1}{q}A\partial _{\theta }+B\partial
_{r}^{2}\right\} \left( 
\begin{array}{l}
N\left( 0,r,\theta \right) \\ 
\Gamma \left( 0,r,\theta \right)%
\end{array}%
\right)  \label{2.1a}
\end{equation}%
$A$ and $B$ being the matrices%
\begin{equation*}
A=\left( 
\begin{array}{ll}
0 & 1 \\ 
1 & 0%
\end{array}%
\right) ;\hspace{0.3cm}B=\left( 
\begin{array}{ll}
D & 0 \\ 
0 & \nu%
\end{array}%
\right)
\end{equation*}%
With $x\circeq \left( r,\theta \right) $, define a function $F$ such that%
\begin{eqnarray}
F\left( x,i\right) &=&N\left( x\right) \hspace{0.3cm}\text{if}\hspace{0.3cm}%
i=+1  \notag \\
F\left( x,i\right) &=&\Gamma \left( x\right) \hspace{0.3cm}\text{if}\hspace{%
0.3cm}i=-1  \label{2.2}
\end{eqnarray}%
Associated to the equations (\ref{2.1}) there is an operator $\mathcal{O}$%
\begin{equation}
\left( \mathcal{O}F\right) \left( x,i\right) =\frac{1}{2}\left\{ 2D\delta
_{i,1}+2\nu \delta _{i,-1}\right\} \partial _{r}^{2}F\left( x,i\right) -%
\frac{1}{q}\delta _{i,-j}\partial _{\theta }F\left( x,j\right)  \label{2.3}
\end{equation}%
which is the generator of the stochastic process associated to the full
linear part of the equation.

However, for the construction of a stochastic solution to the nonlinear
equation through a probabilistic interpretation of the integral equation, it
is convenient to have a stochastic process that operates in a simple way on
the arguments of the function. Therefore instead of the process associated
to $\mathcal{O}$, only the diffusion associated to the first term in (\ref%
{2.3}) will be used below. It also provides an easier handling of the $%
\partial _{\theta }$ derivative.

\subsection{The $\protect\chi =1$ case}

In the $\chi =1$ case the equation (\ref{2.0}) is linear%
\begin{eqnarray}
\partial _{t}N+\frac{1}{q}\partial _{\theta }\Gamma +\frac{1}{\eta }N
&=&D\partial _{r}^{2}N  \notag \\
\partial _{t}\Gamma +\frac{1}{\eta }\left( \Gamma -\Gamma _{0}\right) &=&\nu
\partial _{r}^{2}\Gamma  \label{2.4}
\end{eqnarray}%
the solution being%
\begin{eqnarray}
\left( 
\begin{array}{l}
N\left( t\right) \\ 
\Gamma \left( t\right)%
\end{array}%
\right) &=&e^{t\left( -\frac{1}{q}C\partial _{\theta }+B\partial _{r}^{2}-%
\frac{1}{\eta }\right) }\left\{ \left( 
\begin{array}{l}
N\left( 0\right) \\ 
\Gamma \left( 0\right)%
\end{array}%
\right) \right.  \notag \\
&&+\left. \int_{0}^{t}d\tau e^{-\tau \left( -\frac{1}{q}C\partial _{\theta
}+B\partial _{r}^{2}-\frac{1}{\eta }\right) }\left( 
\begin{array}{c}
0 \\ 
\frac{\Gamma _{0}}{\eta }%
\end{array}%
\right) \right\}  \label{2.5}
\end{eqnarray}%
with $B$ defined before and $C$ being the matrix%
\begin{equation*}
C=\left( 
\begin{array}{ll}
0 & 1 \\ 
0 & 0%
\end{array}%
\right)
\end{equation*}

\subsection{The nonlinear equations $\left( \protect\chi =0\right) $: A
stochastic solution}

For the nonlinear equations one writes%
\begin{eqnarray}
N\left( t,r,\theta \right) &=&e^{tD\partial _{r}^{2}}N\left( 0,r,\theta
\right) -\frac{1}{q}\int_{0}^{t}d\tau e^{\tau D\partial _{r}^{2}}\partial
_{\theta }\Gamma \left( t-\tau ,r,\theta \right)  \notag \\
\Gamma \left( t,r,\theta \right) &=&e^{t\nu \partial _{r}^{2}}\Gamma \left(
0,r,\theta \right) -\frac{1}{q}\int_{0}^{t}d\tau e^{\tau \nu \partial
_{r}^{2}}\partial _{\theta }\left\{ \frac{\Gamma ^{2}}{N}+N\right\} \left(
t-\tau ,r,\theta \right)  \notag \\
&&  \label{2.6}
\end{eqnarray}%
Denote by $\xi _{s}^{(N)}$ and $\xi _{s}^{(\Gamma )}$ two Brownian motions
in the $r-$coordinate with diffusion coefficients $\sqrt{2D}$ and $\sqrt{%
2\nu }$. Then the equations (\ref{2.6}) may be reinterpreted as defining a
probabilistic processes for which the expectation values are the functions $%
N\left( t,r,\theta \right) $ and $\Gamma \left( t,r,\theta \right) $, that is%
\begin{eqnarray}
N\left( t,r,\theta \right) &=&\mathbb{E}_{\left( t,r,\theta \right) }\left[ p%
\frac{1}{p}N\left( 0,\xi _{t}^{(N)},\theta \right) -\frac{t}{q\left(
1-p\right) }\int_{0}^{t}\frac{1-p}{t}d\tau \partial _{\theta }\Gamma \left(
t-\tau ,\xi _{\tau }^{(N)},\theta \right) \right]  \notag \\
\Gamma \left( t,r,\theta \right) &=&\mathbb{E}_{\left( t,r,\theta \right) }%
\left[ p\frac{1}{p}\Gamma \left( 0,\xi _{t}^{(\Gamma )},\theta \right) -%
\frac{2t}{q\left( 1-p\right) }\int_{0}^{t}\frac{1-p}{t}d\tau \partial
_{\theta }\left\{ \frac{1}{2}\frac{\Gamma ^{2}}{N}+\frac{1}{2}N\right\}
\left( t-\tau ,\xi _{\tau }^{(\Gamma )},\theta \right) \right]  \notag \\
&&  \label{2.7}
\end{eqnarray}%
$\mathbb{E}_{\left( t,r,\theta \right) }$ denotes the expectation value of a
stochastic process started from $\left( t,r,\theta \right) $. The processes
that construct the solution at the point $\left( t,r,\theta \right) $ are
backwards-in-time processes that start from time $t$ and propagate to time
zero. With probability $p$ the processes reach time zero and the
contribution to the expectation value is $\frac{1}{p}N\left( 0,\xi
_{t}^{(N)},\theta \right) $ (or $\frac{1}{p}\Gamma \left( 0,\xi
_{t}^{(\Gamma )},\theta \right) $). With probability $\left( 1-p\right) $
the process is interrupted at a time $\tau $ chosen with uniform probability
in the interval $\left( t,0\right) $. For the process associated to $N$, the
process changes its nature, becomes a $\Gamma $ process and picks up a
factor $-\frac{t}{q\left( 1-p\right) }$. For the case of the process $\Gamma 
$, with probability $\frac{1}{2}$, this process either changes to a $N$
process or branches into a $N$ and a $\Gamma $ process. In both cases it
picks up a factor $-\frac{2t}{q\left( 1-p\right) }$.

Notice that the propagation process acts only on the $r-$coordinate.
Therefore the derivative $\partial _{\theta }$, the square in $\Gamma ^{2}$
and the quotient in $\frac{\Gamma ^{2}}{N}$ may all be treated as operators
which are kept as labels at each branching point. When all the lines of the
process reach time zero, the initial condition is sampled at the arrival $%
r_{0}-$coordinate. This initial condition is not simply a value but a
function of $\theta $ ($\Gamma \left( 0,r_{0},\theta \right) $ or $N\left(
0,r_{0},\theta \right) $). It implies that both the initial condition and
all its derivatives at the argument $\theta $ must be provided. This initial
functions are then backtracked throughout the sample lines, the
multiplicative factors are picked up at each $\tau $ interrupt and the
operators applied whenever a labelled branching point is reached. This
provides the contribution of each sample path to the expectation value.
Figure 1 displays an example of a sample path, where the operators picked up
along the way are denoted by flags.

\begin{figure}[htb]
\begin{center}
\psfig{figure=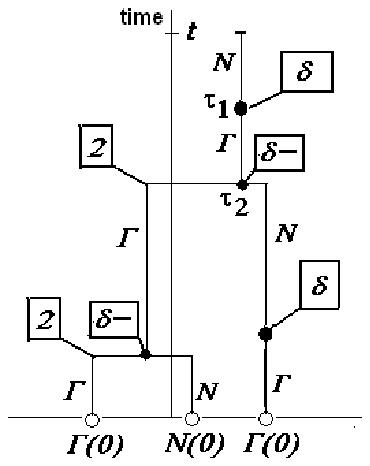,width=11truecm}
\end{center}
\caption{A sample path of the $N-\Gamma$ stochastic process}
\end{figure}

Notice the order of the operators at each branching point. For example, at
the leftmost $\delta -$ labelled point the operation is%
\begin{equation*}
\partial _{\theta }\left\{ \frac{\Gamma ^{2}\left( 0,r_{0}^{(1)},\theta
\right) }{N\left( 0,r_{0}^{(2)},\theta \right) }\right\}
\end{equation*}%
and the whole contribution of this sample path to the $N-$expectation value
is%
\begin{equation*}
\partial _{\theta }^{2}\left\{ \left( \partial _{\theta }\left\{ \frac{%
\Gamma ^{2}\left( 0,r_{0}^{(1)},\theta \right) }{N\left(
0,r_{0}^{(2)},\theta \right) }\right\} \right) ^{2}\left\{ \partial _{\theta
}\Gamma \left( 0,r_{0}^{(3)},\theta \right) \right\} ^{-1}\right\}
\end{equation*}%
times the factor $\left( \frac{1}{p}\right) ^{3}\frac{4t\tau _{1}\tau
_{2}^{2}}{q^{4}\left( 1-p\right) ^{4}}$.

If the initial conditions $\left\vert \frac{\Gamma ^{2}}{N},N,\Gamma
\right\vert $ and all its derivatives are bound by a constant $M$, a worst
case analysis implies that almost sure convergence of the expectation value
is guaranteed for%
\begin{equation*}
\frac{t}{q}M<1
\end{equation*}%
However, in practice, this condition is too severe.

\section{A two-dimensional fluid model for the scrape-off layer (TOKAM 2D)}

Here both the configuration and the Fourier space equations will be analyzed.

\subsection{The configuration space equations}

A two-dimensional fluid model for the scrape-off layer based on the
interchange instability (TOKAM\ 2D \cite{Sarazin}) is%
\begin{equation}
\begin{array}{lll}
\frac{\partial }{\partial t}n & = & S-\{\phi ,n\}-\sigma ne^{\Lambda -\phi
}+D\Delta _{\bot }n \\ 
\frac{\partial }{\partial t}\Delta _{\bot }\phi & = & \sigma \left(
1-e^{\Lambda -\phi }\right) +\nu \Delta _{\bot }^{2}\phi -\{\phi ,\Delta
_{\bot }\phi \}-\frac{1}{n}g\partial _{x_{2}}n%
\end{array}
\label{3.1}
\end{equation}%
where $n=\frac{N}{N_{0}}$ is the normalized density field and $\phi =\frac{eU%
}{T_{e}}$ the normalized electric potential. The brackets $\left\{ \cdots
\right\} $ are Poisson brackets, $\left\{ f,g\right\} =\partial
_{x_{1}}f\partial _{x_{2}}g-\partial _{x_{2}}f\partial _{x_{1}}g$, with $%
x_{1}=\left( r-a\right) /\rho _{s}$ the minor radius normalized by the
Larmor radius $\rho _{s}^{2}=T_{e}/m_{i}$ and $x_{2}=a\theta /\rho _{s}$, $a$
being the plasma radius. $S$ is a source term.

The first equation is rewritten as

\begin{equation}
\begin{array}{lll}
\frac{\partial }{\partial t}n & = & -\sigma e^{\Lambda }n+D\Delta _{\bot
}n+S-\{\phi ,n\}-\sigma e^{\Lambda }\sum_{j=1}\frac{\left( -1\right) ^{j}}{j!%
}n\phi ^{j}%
\end{array}
\label{3.1a}
\end{equation}%
the first two (linear) terms generating the propagating part of the
stochastic process and the others the branching.

The second equation involves $\phi \left( x\right) ,\Delta _{\bot }\phi
\left( x\right) $ and $\Delta _{\bot }^{2}\phi \left( x\right) $. Because
the time derivative acts on $\Delta _{\bot }\phi \left( x\right) $ it is
natural to choose this term as the basic variable. Define%
\begin{equation}
\Psi \left( x\right) =\Delta _{\bot }\phi \left( x\right)  \label{3.2}
\end{equation}%
Then with%
\begin{equation}
K\left( x-y\right) =\frac{1}{2\pi }\log \left\vert x-y\right\vert
\label{3.3a}
\end{equation}%
\begin{equation}
\phi \left( x\right) =\left( K\Psi \right) \left( y\right) =\int
d^{2}yK\left( x-y\right) \Psi \left( y\right)  \label{3.3}
\end{equation}%
the second equation in (\ref{3.1}) becomes

\begin{eqnarray}
\frac{\partial }{\partial t}\Psi &=&-\sigma e^{\Lambda }\Psi +\nu \Delta
_{\bot }\Psi +\sigma \left( 1-e^{\Lambda }\right) +\sigma e^{\Lambda }\left(
K+\delta \right) \Psi  \notag \\
&&-\sigma e^{\Lambda }\sum_{j=2}\frac{\left( -1\right) ^{j}}{j!}\left( K\Psi
\right) ^{j}-\{\left( K\Psi \right) ,\Psi \}-g\partial _{x_{2}}\log n
\label{3.5}
\end{eqnarray}%
where $\delta $ stands for the kernel $\delta \left( x-y\right) $. Notice
that from the negative contribution at $x=y$ of the term $\sigma e^{\Lambda
}K\Psi $ (in the expansion of $-\sigma e^{-\phi }$), a negative part is
extracted which becomes a local dissipative term.

The integral version of the equations (\ref{3.1a}) and (\ref{3.5}) is%
\begin{eqnarray}
n\left( x,t\right)  &=&e^{-t\sigma e^{\Lambda }}e^{tD\Delta _{\bot }}n\left(
x,0\right)   \notag \\
&&+\int_{0}^{t}dse^{-s\sigma e^{\Lambda }}e^{sD\Delta _{\bot }}\left\{
S-\{K\Psi ,n\}-\sigma e^{\Lambda }\sum_{j=1}\frac{\left( -1\right) ^{j}}{j!}%
n\left( K\Psi \right) ^{j}\right\} \left( x,t-s\right)   \notag \\
&&  \label{3.6}
\end{eqnarray}%
\begin{eqnarray}
\Psi \left( x,t\right)  &=&e^{-t\sigma e^{\Lambda }}e^{t\nu \Delta _{\bot
}}\Psi \left( x,0\right) +\int_{0}^{t}e^{-s\sigma e^{\Lambda }}e^{s\nu
\Delta _{\bot }}\left\{ \sigma \left( 1-e^{\Lambda }\right) +\sigma
e^{\Lambda }\left( K+\delta \right) \Psi \right.   \notag \\
&&\left. -\sigma e^{\Lambda }\sum_{j=2}\frac{\left( -1\right) ^{j}}{j!}%
\left( K\Psi \right) ^{j}-\{\left( K\Psi \right) ,\Psi \}-g\partial
_{x_{2}}\log n\right\} \left( x,t-s\right)   \label{3.7}
\end{eqnarray}%
The integral equations will now be given a probabilistic interpretation. In
addition to the usual propagation and branching mechanisms, the kernel
integrations $\left( K\Psi \right) \left( x\right) =\int d^{2}yK\left(
x-y\right) \Psi \left( y\right) $ and $\left( \partial _{x_{i}}K\Psi \right)
\left( x\right) $ must also be given a probabilistic interpretation. For
this purpose two questions have to be dealt with. First, the kernel
integrals $\int d^{2}yK\left( x-y\right) $ and $\int d^{2}y\partial
_{x_{i}}K\left( x-y\right) $ are not finite, second $K\left( x-y\right) $
and $\partial _{x_{i}}K\left( x-y\right) $ are not positive definite. A
positive function $h\left( y\right) $ is chosen in such a way that%
\begin{equation*}
N\left( x\right) =\int d^{2}y\left\vert K\left( x-y\right) \right\vert
h\left( y\right) ;N_{\delta }\left( x\right) =\int d^{2}y\left\vert K+\delta
\right\vert h\left( y\right) ;N_{i}\left( x\right) =\int d^{2}y\left\vert
\partial _{x_{i}}K\right\vert h\left( y\right) 
\end{equation*}%
are finite\footnote{%
There are many functions satisfying this requirement. For example $h\left(
y\right) =\left\vert y\right\vert e^{-\left\vert y\right\vert }$}. Then%
\begin{eqnarray*}
\rho \left( x,y\right)  &=&\frac{\left\vert K\left( x-y\right) \right\vert
h\left( y\right) }{N\left( x\right) };\rho _{\delta }\left( x,y\right) =%
\frac{\left\vert K\left( x-y\right) +\delta \left( x-y\right) \right\vert
h\left( y\right) }{N_{\delta }\left( x\right) }; \\
\rho _{i}\left( x,y\right)  &=&\frac{\left\vert \partial _{i}K\left(
x-y\right) \right\vert h\left( y\right) }{N_{i}\left( x\right) }
\end{eqnarray*}%
may be considered as $x-$dependent probability densities in $y\in \mathbb{R}%
^{2}$. Define%
\begin{equation*}
\Omega \left( x,t\right) =\frac{\Psi \left( x,t\right) }{h\left( x\right) }
\end{equation*}%
Then, Eqs.(\ref{3.6}) and (\ref{3.7}) are rewritten%
\begin{eqnarray}
n\left( x,t\right)  &=&e^{-t\sigma e^{\Lambda }}e^{tD\Delta _{\bot }}n\left(
x,0\right) +\int_{0}^{t}ds\sigma e^{\Lambda }e^{-s\sigma e^{\Lambda
}}e^{sD\Delta _{\bot }}\left\{ p_{s,n}M_{s,n}S\left( x\right) \right.  
\notag \\
&&\left. +p_{B,n}^{(1)}M_{2,n}\partial _{x_{2}}n\int d^{2}y\rho _{1}\left(
x,y\right) \Omega \left( y\right) +p_{B,n}^{(2)}M_{1,n}\partial
_{x_{1}}n\int d^{2}y\rho _{2}\left( x,y\right) \Omega \left( y\right)
\right.   \notag \\
&&\left. +p_{\Sigma ,n}\sum_{j=1}p_{j,n}M_{j,n}n\left( x\right) \left( \int
d^{2}y\rho \left( x,y\right) \Omega \left( y\right) \right) ^{j}\right\}
\left( t-s\right)   \label{3.8}
\end{eqnarray}%
\begin{eqnarray}
\Omega \left( x,t\right)  &=&e^{-t\sigma e^{\Lambda }}e^{t\nu \Delta _{\bot
}}\Omega \left( x,0\right) +\int_{0}^{t}ds\sigma e^{\Lambda }e^{-s\sigma
e^{\Lambda }}e^{s\nu \Delta _{\bot }}\left\{ p_{k}M_{k}+p_{\delta }M_{\delta
}\int d^{2}y\rho _{\delta }\left( x,y\right) \Omega \left( y\right) \right. 
\notag \\
&&\left. +p_{\Sigma ,\Omega }\sum_{j=2}p_{j,\Omega }M_{j,\Omega }\left( \int
d^{2}y\rho \left( x,y\right) \Omega \left( y\right) \right) ^{j}\right.  
\notag \\
&&\left. +p_{B,\Omega }^{(1)}\left( \frac{1}{2}M_{2,\Omega }^{(a)}\partial
_{x_{2}}\Omega +\frac{1}{2}M_{2,\Omega }^{(b)}\Omega \right) \int d^{2}y\rho
_{1}\left( x,y\right) \Omega \left( y\right) \right.   \notag \\
&&\left. +p_{B,\Omega }^{(2)}\left( \frac{1}{2}M_{1,\Omega }^{(a)}\partial
_{x_{1}}\Omega +\frac{1}{2}M_{1,\Omega }^{(b)}\Omega \right) \int d^{2}y\rho
_{2}\left( x,y\right) \Omega \left( y\right) +p_{g}M_{g}\partial
_{x_{2}}\log n\right\} \left( t-s\right)   \notag \\
&&  \label{3.9}
\end{eqnarray}%
with probabilities%
\begin{eqnarray}
p_{s,n} &=&\left( 3+\sigma e^{\Lambda }\right) ^{-1}  \notag \\
p_{B,n}^{(1)} &=&p_{B,n}^{(2)}=\left( 3+\sigma e^{\Lambda }\right) ^{-1} 
\notag \\
p_{\Sigma ,n} &=&\sigma e^{\Lambda }\left( 3+\sigma e^{\Lambda }\right) ^{-1}
\notag \\
p_{j,n} &=&\frac{1}{j!\left( e-1\right) }  \notag \\
p_{k} &=&\left( 2+\sigma +2\sigma e^{\Lambda }+g\right) ^{-1}  \notag \\
p_{B,\Omega }^{(1)} &=&p_{B,\Omega }^{(2)}=\left( 2+\sigma +2\sigma
e^{\Lambda }+g\right) ^{-1}  \notag \\
p_{\delta } &=&\sigma e^{\Lambda }\left( 2+\sigma +2\sigma e^{\Lambda
}+g\right) ^{-1}  \notag \\
p_{\Sigma ,\Omega } &=&\sigma e^{\Lambda }\left( 2+\sigma +2\sigma
e^{\Lambda }+g\right) ^{-1}  \notag \\
p_{g} &=&g\left( 2+\sigma +2\sigma e^{\Lambda }+g\right) ^{-1}  \notag \\
p_{j,\Omega } &=&\frac{1}{j!\left( e-2\right) }  \label{3.10}
\end{eqnarray}%
and multipliers%
\begin{eqnarray}
M_{s,n} &=&1+3/\left( \sigma e^{\Lambda }\right)   \notag \\
M_{2,n} &=&-N_{1}\left( x\right) \left( 1+3/\left( \sigma e^{\Lambda
}\right) \right)   \notag \\
M_{1,n} &=&N_{2}\left( x\right) \left( 1+3/\left( \sigma e^{\Lambda }\right)
\right)   \notag \\
M_{j,n} &=&-\left( -1\right) ^{j}\left( e-1\right) \left( 1+3/\left( \sigma
e^{\Lambda }\right) \right) N\left( x\right) ^{j}  \notag \\
M_{k} &=&\left( 1-e^{\Lambda }\right) \left( 2+\sigma +2\sigma e^{\Lambda
}+g\right) \left( h\left( x\right) \sigma e^{\Lambda }\right) ^{-1}  \notag
\\
M_{2,\Omega }^{(a)} &=&-2\left( 2+\sigma +2\sigma e^{\Lambda }+g\right)
N_{1}\left( x\right) \left( \sigma e^{\Lambda }\right) ^{-1}  \notag \\
M_{1,\Omega }^{(a)} &=&2\left( 2+\sigma +2\sigma e^{\Lambda }+g\right)
N_{2}\left( x\right) \left( \sigma e^{\Lambda }\right) ^{-1}  \notag \\
M_{2,\Omega }^{(b)} &=&-2\left( 2+\sigma +2\sigma e^{\Lambda }+g\right)
N_{1}\left( x\right) \left( \sigma e^{\Lambda }\right) ^{-1}\partial
_{x_{2}}\log h\left( x\right)   \notag \\
M_{1,\Omega }^{(b)} &=&2\left( 2+\sigma +2\sigma e^{\Lambda }+g\right)
N_{2}\left( x\right) \left( \sigma e^{\Lambda }\right) ^{-1}\partial
_{x_{1}}\log h\left( x\right)   \notag \\
M_{\delta } &=&N_{\delta }\left( x\right) \left( 2+\sigma +2\sigma
e^{\Lambda }+g\right) \left( h\left( x\right) \sigma e^{\Lambda }\right)
^{-1}  \notag \\
M_{g} &=&-\left( 2+\sigma +2\sigma e^{\Lambda }+g\right) \left( h\left(
x\right) \sigma e^{\Lambda }\right) ^{-1}  \notag \\
M_{j,\Omega } &=&-\left( -1\right) ^{j}\left( e-2\right) \left( h\left(
x\right) \sigma e^{\Lambda }\right) ^{-1}\left( 2+\sigma +2\sigma e^{\Lambda
}+g\right) N\left( x\right) ^{j}  \label{3.11}
\end{eqnarray}

In this form the integral equations may be solved by two stochastic process,
for $n$ and $\Omega $, which starting from time $t$, propagate
backwards-in-time as Brownian motions with diffusion coefficients $D$ and $%
\nu $, respectively. These processes either reach time zero with probability 
$e^{-t\sigma e^{\Lambda }}$ or branch at time $t-s$ with probability density 
$\sigma e^{\Lambda }e^{-s\sigma e^{\Lambda }}$. At each branching point the
appropriate branching is chosen according to the probabilities $p_{s,n}$, $%
p_{B,n}^{(1)}$, $p_{B,n}^{(2)}$, $p_{\Sigma ,n}$ for the $n-$process and $%
p_{k}$, $p_{B,\Omega }^{(1)}$, $p_{B,\Omega }^{(2)}$, $p_{\delta }$, $%
p_{\Sigma ,\Omega }$, $p_{g}$ for the $\Omega -$process. With probability $%
p_{s,n}$ the $n-$process samples the source $S$ and with probability $p_{k}$
the $\Omega -$process is killed. When it is $\Sigma ,n$ or $\Sigma ,\Omega $
that is chosen, the branching has $j$ branches according to the probability $%
p_{j,n}$ or $p_{j,\Omega }$. For the branches of the $n-$type the argument $%
x $ becomes $x^{\prime }=e^{sD\Delta _{\bot }}x$ or $x^{\prime }=e^{s\nu
\Delta _{\bot }}x$, that is, it is propagated by the Brownian motions. For
the branches of $\Omega -$type the argument $y$ is chosen with $x-$dependent
probability densities $\rho (x,y),\rho _{\delta }\left( x,y\right) $ or $%
\rho _{i}\left( x,y\right) $. At each branching point the multipliers in (%
\ref{3.11}) are picked up as multiplicative factors. For the $\Omega -$%
process, the branching associated to the Poisson bracket term has two
possibilities occurring with probability $\frac{1}{2}$ each. They are either 
$\partial _{x_{i}}\Omega \left( x\right) \Omega \left( y\right) $ or $\Omega
\left( x\right) \Omega \left( y\right) $ with multipliers $M_{i,\Omega
}^{(a)}$ or $M_{i,\Omega }^{(b)}$. These two branchings originate from the
term%
\begin{equation*}
\frac{1}{h\left( x\right) }\partial _{x_{i}}\left( h\left( x\right) \Omega
\left( x\right) \right) =\Omega \left( x\right) \partial _{x_{i}}\log
h\left( x\right) +\partial _{x_{i}}\Omega \left( x\right)
\end{equation*}%
This splitting is essential because the operator $\frac{1}{h\left( x\right) }%
\partial _{x_{i}}h\left( x\right) $ does not commute with the time evolution 
$e^{t\nu \Delta _{\bot }}$.

The final step in the specification of how the stochastic processes lead to
a solution of the equations (\ref{3.6}) and (\ref{3.7}) is the handling of
the derivatives $\partial _{x_{i}}n\left( x\right) $ and $\partial
_{x_{i}}\Omega \left( x\right) $. Contrary to the case of the SOLEDGE
equation treated in Section 2, the derivatives in this case act on the same
variables as the stochastic process. However, $\partial _{x_{1}}$ and $%
\partial _{x_{2}}$ commute with the time evolution operators $e^{tD\Delta
_{\bot }}$ or $e^{t\nu \Delta _{\bot }}$. Hence they may be kept as operator
labels at the branching points and proceed with the evolution of the fields $%
n$ and $\Omega $. When the fields finally reach time zero, the calculation
is made by backtracking (forward in time) through the tree the values of the
fields and their derivatives at the final points and performing the
operations at each labelled vertex. Notice that the operators are kept at
the vertices and not carried along by the fields, because at non linear
vertices the Leibnitz rule should be applied. For example, suppose that from
a vertex $\partial _{i}\Omega \left( x_{1}\right) $ at time $s_{1}$ the
field later branches into $\Omega \left( x_{2}\right) \Omega \left(
y_{2}\right) $ which then reach time zero at the points $x_{0}^{(1)}$ and $%
x_{0}^{(2)}$. Then the contribution of this derivative vertex is%
\begin{equation*}
\Omega \left( x_{0}^{(1)}\right) \partial _{i}\Omega \left(
x_{0}^{(2)}\right) +\partial _{i}\Omega \left( x_{0}^{(1)}\right) \Omega
\left( x_{0}^{(2)}\right)
\end{equation*}%
The derivatives are computed at the arrival points, whereas the terms $%
h\left( x\right) $ and $\log h\left( x\right) $ in the multipliers (which do
not commute with time evolution) are computed at the branching points. That
this is the correct procedure is easily understood by recursively iterating
the equations. Figure 2 shows how a sampled path of these processes looks
like.

\begin{figure}[htb]
\begin{center}
\psfig{figure=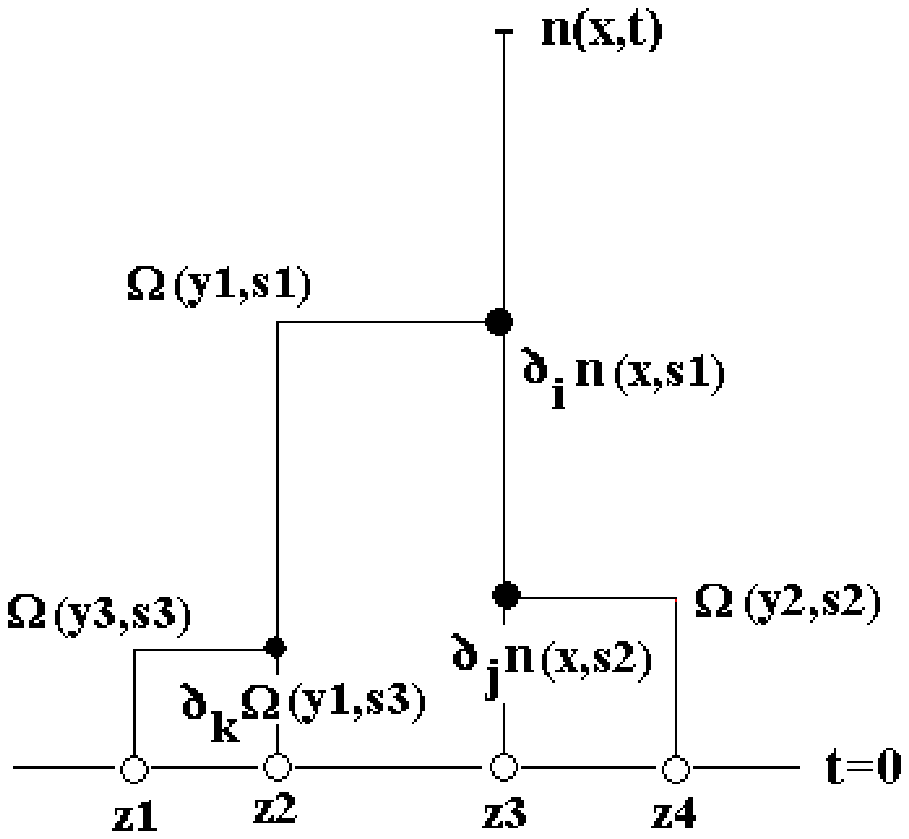,width=11truecm}
\end{center}
\caption{A sample path of the $n-$process}
\end{figure}

The contribution of this sample path to the $n-$expectation value is%
\begin{equation*}
M_{i,n}\left( x\right) M_{j,n}\left( x\right) \left\{ \partial
_{j}n_{0}\left( z_{3}\right) \partial _{i}\Omega _{0}\left( z_{4}\right)
+\partial _{i}\partial _{j}n_{0}\left( z_{3}\right) \Omega _{0}\left(
z_{4}\right) \right\} M_{k,\Omega }^{(a)}\left( y_{1}\right) \Omega
_{0}\left( z_{1}\right) \partial _{k}\Omega _{0}\left( z_{2}\right)
\end{equation*}

Some remarks on this construction:

(i) The normalization of the kernels by the $h\left( x\right) $ function
converts the computation of the integral into a probabilistic sampling. At
the same time the kernel itself absorbs one of the derivatives, that
otherwise would act on the fields.

(ii) The exponential $e^{-\phi }$ was expanded into a power series leading
to multiple branchings with probabilities $p_{j,n}$ and $p_{j,\Omega }$. By
contrast the other nonpolynomial term $\log n$ was kept as an operator. The
reason is that the coefficients in the series expansion of the $\log $ do
not lead to a probability distribution.

(iii) The complexity of the resulting processes, reflects the complexity of
the original equations (\ref{3.1}). Nevertheless what this construction
shows is that clearly defined techniques may be used to construct stochastic
solutions even when derivatives and complex non-linear terms are present.

To insure convergence of the calculation of the solution by the stochastic
processes, a bound $M$ must be put on the multipliers (which depend on the
normalization factors, hence on the $h\left( x\right) $ function), on the
initial conditions $n_{0}\left( x\right) $ and $\Omega _{0}\left( x\right) $
and their derivatives, on the source term $S$ and on $\frac{\sigma \left(
1-e^{\Lambda }\right) }{h\left( x\right) }$. From the branching process one
estimates the probability of a tree of $n$ branches, therefore%
\begin{equation*}
M\leq \frac{1}{1-e^{-\sigma e^{\Lambda }t}}
\end{equation*}

\subsection{The Fourier-transformed TOKAM 2D}

In the Fourier transformed equation, the derivatives become simpler
multiplicative factors. However the nonlinear terms become more complex. Let%
\begin{equation*}
f=\log n
\end{equation*}%
Then, the first equation in (\ref{3.1}) becomes%
\begin{equation}
\frac{\partial }{\partial t}f=S^{\prime }-\{\phi ,f\}-\sigma e^{\Lambda
-\phi }+D\Delta _{\bot }f+D\left\vert \nabla f\right\vert ^{2}  \label{4.1}
\end{equation}%
where the source term has been redefined as $S^{\prime }=\frac{S}{n}$. Let
the Fourier transforms of $f$ and $\phi $ be%
\begin{eqnarray*}
F\left( k,t\right) &=&\frac{1}{2\pi }\int d^{2}xf\left( x,t\right)
e^{ik\cdot x} \\
\Phi \left( k,t\right) &=&\frac{1}{2\pi }\int d^{2}x\phi \left( x,t\right)
e^{ik\cdot x}
\end{eqnarray*}%
Then, the Fourier transform of the equations (\ref{3.1}) is%
\begin{eqnarray}
\frac{\partial }{\partial t}F\left( k,t\right) &=&-D\left\vert k\right\vert
^{2}F\left( k,t\right) +\overset{\backsim }{S^{\prime }}\left( k,t\right) +%
\frac{1}{2\pi }\int d^{2}\xi \left( k_{1}-\xi _{1}\right) \xi _{2}\left\{
\Phi \left( k-\xi \right) F\left( \xi \right) \right.  \notag \\
&&\left. -F\left( k-\xi \right) \Phi \left( \xi \right) \right\}  \notag \\
&&-\sigma e^{\Lambda }\sum_{n=0}^{\infty }\frac{\left( -1\right) ^{n}}{%
n!\left( 2\pi \right) ^{n-1}}\Phi ^{\ast n}\left( k\right) -\frac{D}{2\pi }%
\int d^{2}\xi \left\{ \left( k_{1}-\xi _{1}\right) \xi _{1}\right.  \notag \\
&&+\left. \left( k_{2}-\xi _{2}\right) \xi _{2}\right\} F\left( k-\xi
\right) F\left( \xi \right)  \notag \\
\frac{\partial }{\partial t}\Phi \left( k,t\right) &=&-\nu \left\vert
k\right\vert ^{2}\Phi \left( k,t\right) +\frac{1}{2\pi }\int \frac{d^{2}\xi 
}{\left\vert k\right\vert ^{2}}\left( k_{1}-\xi _{1}\right) \xi _{2}\Phi
\left( k-\xi \right) \Phi \left( \xi \right) \left\{ \left\vert \xi
\right\vert ^{2}-\left\vert k-\xi \right\vert ^{2}\right\}  \notag \\
&&-ig\frac{k_{2}}{\left\vert k\right\vert ^{2}}F\left( k\right) -\frac{%
\sigma }{\left\vert k\right\vert ^{2}}\left\{ 2\pi \delta ^{2}\left(
k\right) -e^{\Lambda }\sum_{n=0}^{\infty }\frac{\left( -1\right) ^{n}}{%
n!\left( 2\pi \right) ^{n-1}}\Phi ^{\ast n}\left( k\right) \right\}
\label{4.2}
\end{eqnarray}%
where $\overset{\backsim }{S^{\prime }}$ is the Fourier transform of $%
S^{\prime }$ and $\Phi ^{\ast n}$ denotes the convolution power.%
\begin{equation*}
\Phi ^{\ast n}\left( k\right) =\int \prod_{i=1}^{n}d^{2}\xi ^{(i)}\delta
\left( k-\sum_{l=1}^{n}\xi ^{(l)}\right) \prod_{p=1}^{n}\Phi \left( \xi
^{(p)}\right)
\end{equation*}

To control the growth of the functional obtained by the stochastic process
one divides $F\left( k,t\right) $ and $\Phi \left( k,t\right) $ by a
majorizing kernel $\gamma \left( k\right) $%
\begin{eqnarray}
\chi \left( k,t\right) &=&\frac{F\left( k,t\right) }{\gamma \left( k\right) }
\notag \\
\zeta \left( k,t\right) &=&\frac{\Phi \left( k,t\right) }{\gamma \left(
k\right) }  \label{4.3}
\end{eqnarray}%
and writes integral equations for $\chi \left( k,t\right) $%
\begin{eqnarray}
\chi \left( k,t\right) &=&e^{-D\left\vert k\right\vert ^{2}t}\chi \left(
k,0\right) +\int_{0}^{t}dsD\left\vert k\right\vert ^{2}e^{-D\left\vert
k\right\vert ^{2}s}\left\{ p_{a}\int d^{2}\xi M_{a}\left( k,\xi \right)
p\left( k,\xi \right) \right.  \notag \\
&&\left[ \frac{1}{2}\zeta \left( k-\xi ,t-s\right) \chi \left( \xi
,t-s\right) -\frac{1}{2}\chi \left( k-\xi ,t-s\right) \zeta \left( \xi
,t-s\right) \right]  \notag \\
&&+p_{b}\sum_{n=0}p_{n}\int d^{2}\xi _{1}\cdots d^{2}\xi _{n-1}M_{n}\left(
k,\xi _{1},\cdots ,\xi _{n-1}\right) p\left( k,\xi _{1},\cdots ,\xi
_{n-1}\right)  \notag \\
&&\times \zeta \left( k-\sum_{i=1}^{n-1}\xi _{i},t-s\right) \zeta \left( \xi
_{1},t-s\right) \cdots \zeta \left( \xi _{n-1},t-s\right)  \notag \\
&&+p_{c}\int d^{2}\xi p\left( k,\xi \right) \left[ \frac{1}{2}%
M_{c}^{(1)}\left( k,\xi \right) +\frac{1}{2}M_{c}^{(2)}\left( k,\xi \right) %
\right] \chi \left( k-\xi ,t-s\right) \chi \left( \xi ,t-s\right)  \notag \\
&&\left. +p_{s}M_{s}\left( k\right) \frac{\overset{\backsim }{S^{\prime }}%
\left( k,t-s\right) }{\gamma \left( k\right) }\right\}  \label{4.4}
\end{eqnarray}%
with probabilities%
\begin{eqnarray}
p_{a} &=&\left( 1+D+2\pi +2\pi \sigma e^{\Lambda }\right) ^{-1}  \notag \\
p_{b} &=&2\pi \sigma e^{\Lambda +1}\left( 1+D+2\pi +2\pi \sigma e^{\Lambda
}\right) ^{-1}  \notag \\
p_{c} &=&D\left( 1+D+2\pi +2\pi \sigma e^{\Lambda }\right) ^{-1}  \notag \\
p_{s} &=&2\pi \left( 1+D+2\pi +2\pi \sigma e^{\Lambda }\right) ^{-1}  \notag
\\
p_{n} &=&\left( en!\right) ^{-1}  \notag \\
p\left( k,\xi _{1},\cdots ,\xi _{n-1}\right) &=&\frac{\gamma \left(
k-\sum_{i=1}^{n-1}\xi _{i}\right) \gamma \left( \xi _{1}\right) \cdots
\gamma \left( \xi _{n-1}\right) }{\gamma ^{\ast n}\left( k\right) }
\label{4.5}
\end{eqnarray}%
and multipliers%
\begin{eqnarray}
M_{a}\left( k,\xi \right) &=&\frac{\left( k_{1}-\xi _{1}\right) \xi
_{2}\gamma \ast \gamma \left( k\right) }{\gamma \left( k\right) D\left\vert
k\right\vert ^{2}\pi }\left( 1+D+2\pi +2\pi \sigma e^{\Lambda }\right) 
\notag \\
M_{c}^{(1)}\left( k,\xi \right) &=&-\frac{\left( k_{1}-\xi _{1}\right) \xi
_{1}\gamma \ast \gamma \left( k\right) }{\gamma \left( k^{2}\right)
\left\vert k\right\vert ^{2}\pi }\left( 1+D+2\pi +2\pi \sigma e^{\Lambda
}\right)  \notag \\
M_{c}^{(2)}\left( k,\xi \right) &=&-\frac{\left( k_{2}-\xi _{2}\right) \xi
_{2}\gamma \ast \gamma \left( k\right) }{\gamma \left( k^{2}\right)
D\left\vert k\right\vert ^{2}\pi }\left( 1+D+2\pi +2\pi \sigma e^{\Lambda
}\right)  \notag \\
M_{n}\left( k\right) &=&-\frac{\left( -1\right) ^{n}e\gamma ^{\ast n}\left(
k\right) }{\left( 2\pi \right) ^{n}\gamma \left( k\right) D\left\vert
k\right\vert ^{2}}\left( 1+D+2\pi +2\pi \sigma e^{\Lambda }\right)  \notag \\
M_{s}\left( k\right) &=&\frac{1}{2\pi D\left\vert k\right\vert ^{2}}\left(
1+D+2\pi +2\pi \sigma e^{\Lambda }\right)  \label{4.6}
\end{eqnarray}%
Notice that the term $n=0$ in the sum is $p_{b}p_{0}M_{0}\delta \left(
k^{2}\right) $ which, when chosen, kills the contribution of the
corresponding sample path for $k\neq 0$.

For $\varsigma \left( k,t\right) $%
\begin{eqnarray}
\zeta \left( k,t\right) &=&e^{-\nu \left\vert k\right\vert ^{2}t}\zeta
\left( k,0\right) +\int_{0}^{t}ds\nu \left\vert k\right\vert ^{2}e^{-\nu
\left\vert k\right\vert ^{2}s}\left\{ p_{a}^{\prime }\int d^{2}\xi
M_{a}^{\prime }\left( k,\xi \right) p\left( k,\xi \right) \right.  \notag \\
&&\times \zeta \left( k-\xi ,t-s\right) \zeta \left( \xi ,t-s\right)
+p_{b}^{\prime }\sum p_{n}\int d^{2}\xi _{1}\cdots d^{2}\xi
_{n-1}M_{n}^{\prime }\left( k,\xi _{1},\cdots ,\xi _{n-1}\right)  \notag \\
&&\times p\left( k,\xi _{1},\cdots ,\xi _{n-1}\right) \zeta \left(
k-\sum_{i=1}^{n-1}\xi _{i},t-s\right) \cdots \zeta \left( \xi
_{n-1},t-s\right)  \notag \\
&&\left. +p_{\chi }M_{\chi }\left( k,\xi \right) \chi \left( k,t-s\right)
\right\}  \label{4.7}
\end{eqnarray}%
with probabilities and multipliers%
\begin{eqnarray}
p_{a}^{\prime } &=&\left( \frac{1}{2\pi }+g+\sigma \right) ^{-1}\frac{1}{%
2\pi }  \notag \\
p_{b}^{\prime } &=&\left( \frac{1}{2\pi }+g+\sigma \right) ^{-1}\sigma 
\notag \\
p_{\chi } &=&\left( \frac{1}{2\pi }+g+\sigma \right) ^{-1}g  \notag \\
p_{n} &=&\left( en!\right) ^{-1}  \notag \\
p\left( k,\xi _{1},\cdots ,\xi _{n-1}\right) &=&\frac{\gamma \left(
k-\sum_{i=1}^{n-1}\xi _{i}\right) \gamma \left( \xi _{1}\right) \cdots
\gamma \left( \xi _{n-1}\right) }{\gamma ^{\ast n}\left( k\right) }
\label{4.8}
\end{eqnarray}%
\begin{eqnarray}
M_{a}^{\prime }\left( k,\xi \right) &=&\frac{\left( k_{1}-\xi _{1}\right)
\xi _{2}\left\{ \left\vert \xi \right\vert ^{2}-\left\vert k-\xi \right\vert
^{2}\right\} \gamma \ast \gamma \left( k\right) }{\gamma \left( k\right) \nu
\left\vert k\right\vert ^{4}}\left( \frac{1}{2\pi }+g+\sigma \right)  \notag
\\
M_{n}\left( k\right) &=&\frac{\left( -1\right) ^{n}e^{\Lambda }\gamma ^{\ast
n}\left( k\right) }{\left( 2\pi \right) ^{n-1}\gamma \left( k\right) \nu
\left\vert k\right\vert ^{4}}\left( \frac{1}{2\pi }+g+\sigma \right) \hspace{%
0.5cm}n\geq 1  \notag \\
M_{0}\left( k\right) &=&\frac{2\pi \left( e^{\Lambda }-1\right) }{\gamma
\left( k\right) \nu \left\vert k\right\vert ^{4}}\left( \frac{1}{2\pi }%
+g+\sigma \right)  \notag \\
M_{\chi }\left( k\right) &=&\frac{-ik_{2}}{\nu \left\vert k\right\vert ^{4}}%
\left( \frac{1}{2\pi }+g+\sigma \right)  \label{4.9}
\end{eqnarray}%
The probabilistic interpretation of these equations is similar to the
previous cases. There are two stochastic processes that started at $t$,
propagate backwards in time as Brownian motions with coefficients $%
D\left\vert k\right\vert ^{2}$ and $\nu \left\vert k\right\vert ^{2}$. With
probabilities $e^{-D\left\vert k\right\vert ^{2}t}$ or $e^{-\nu \left\vert
k\right\vert ^{2}t}$ the processes either reach $t=0$ without branching or
branch with probability densities $D\left\vert k\right\vert
^{2}e^{-D\left\vert k\right\vert ^{2}s}$ or $\nu \left\vert k\right\vert
^{2}e^{-\nu \left\vert k\right\vert ^{2}s}$ at time $t-s$. The branch that
they follow afterwards is controlled by the probabilities in (\ref{4.5}) and
(\ref{4.8}). Because products become convolutions under the Fourier
transform, probability densities are needed to decide the new momenta of the
branches. The main role of the majorizing kernel $\gamma \left( k\right) $
is to normalize these probability densities in momentum space. At each
branching point the process picks up the multipliers in (\ref{4.6}) and (\ref%
{4.9}), the contribution of each sample path to the the expectation values
of $\chi \left( k,t\right) $ and $\zeta \left( k,t\right) $ being the
initial values of the corresponding fields $\chi \left( k,0\right) $ or $%
\zeta \left( k,0\right) $ when the processes reach time zero multiplied by
the multipliers picked up along the path. In this case because there are no
operator labels, there is no need to backtrack forward in time as in the
configuration space equation. The price one pays is a more complex branching
structure at each vertex.

A strict condition for the convergence of the process is obtained by a bound 
$M\leq 1$ on all the factors that intervene in the final calculation of each
path contribution.

\section{Remarks and conclusions}

1) The scrape-off layer equations provide a good testing ground for the
construction of stochastic solutions when the equations contain
nonpolynomial terms and derivatives. Therefore, the techniques developed
here considerably extend the range of equations to which the stochastic
solution technique may be applied.

2) To obtain good accuracy in the stochastic solutions, many sample paths
should be computed for each starting configuration. How many samples are
needed may be estimated by large deviation techniques \cite{Elena}. The
sample independence character of the stochastic calculation reduces the
severity of this problem when using parallel computing. Nevertheless it
should be pointed out that, when solutions are desired over a large domain,
the stochastic method is computationally competitive only when applied
together with the domain decomposition method \cite{Acebron1} \cite{Acebron2}
\cite{Acebron3}. Then a considerable improvement is obtained. However, if
local solutions on configuration or Fourier space are desired, the
stochastic method is quite appropriate. For example a study of the time
evolution of a few high Fourier modes gives information on the turbulence
spectrum that only a very fine grid and an expensive calculation would
provide with a global deterministic algorithm.

\end{document}